\documentclass[aps, prl,superscriptaddress, twocolumn,showpacs,preprintnumbers,amsmath,amssymb]{revtex4}

\usepackage{graphicx}
\usepackage{dcolumn}
\usepackage{bm}
\usepackage{amsmath}
\usepackage{amssymb}
\usepackage[ansinew]{inputenc}

\newcommand{\ANGS}{\msz \textrm{\AA/s}} 

\newcommand{\GEMN}{$\textrm{Ge}_{0.95}\textrm{Mn}_{0.05}$}
\newcommand{\GRADCM}{^\circ\textrm{C}}
\newcommand{\GRADKM}{\msz \textrm{K}}
\newcommand{\MF}[1]{$\mu_0 \textrm{H} = #1\msz \textrm{T}$}
\newcommand{\msz}{\,} 

\newcommand{\NM}[1]{#1\msz\textrm{nm}}

\newcommand{\TCC}{$T_\textrm C ^\textrm{Cluster}$}

\newcommand{\TS}[1]{$T_\textrm S = #1\GRADCM$}
\newcommand{\TSS}{$T_\textrm S$}
\begin{document}

\preprint{~}

\title{Clustering in a precipitate free GeMn magnetic semiconductor}

\author{D. Bougeard}
\author{S. Ahlers}
    \affiliation{Walter Schottky Institut, Technische Universität München, Am Coulombwall 3, D-85748 Garching, Germany}
\author{A. Trampert}
    \affiliation{Paul-Drude-Institut für Festkörperelektronik, Hausvogteiplatz 5-7, D-10117 Berlin, Germany}
\author{N. Sircar}
\author{G. Abstreiter}
  \affiliation{Walter Schottky Institut, Technische Universität München, Am Coulombwall 3, D-85748 Garching, Germany}

\date{\today}

\begin{abstract}
We present the first study relating structural parameters of
precipitate free $\textrm{Ge}_{0.95}\textrm{Mn}_{0.05}$ films to magnetisation data. Nanometer sized
clusters - areas with increased Mn content on substitutional lattice
sites compared to the host matrix - are detected in transmission
electron microscopy (TEM) analysis. The films show no overall
spontaneous magnetisation at all down to $2\, \textrm{K}$. The TEM and
magnetisation results are interpreted in terms of an assembly of
superparamagnetic moments developing in the dense distribution of
clusters. Each cluster individually turns ferromagnetic below an
ordering temperature which depends on its volume and Mn content.
\end{abstract}

\maketitle

The development of a novel class of materials combining standard
semiconductors with magnetic elements has recently been driven by
considerable technological as well as fundamental scientific
interest. While the possibility of a seamless combination of
magnetic and semiconducting systems using spins as an additional degree of
freedom opens stimulating perspectives in the field of electronics \cite{wolf:2006IJRD,tanaka:2005JCG}, reports on materials
displaying both semiconducting and ferromagnetic properties have
induced great theoretical and experimental efforts in the
understanding of the underlying physics \cite{dietl:2000S}. Ga(Mn)As
today represents one of the best understood ferromagnets. This
material is one example of a diluted magnetic semiconductor (DMS),
meaning a dispersion of the magnetic elements without affecting the
semiconducting properties of the matrix \cite{ohno:1998S}. The
realisation of DMS with maximised ferromagnetic ordering
temperatures $T_\textrm C$ represents the ultimate objective in this
field.

Special attention has been given to technologically important group
IV semiconductor based magnetic materials, with a prominent position
for GeMn. Since the first claim of the realisation of a Ge based DMS
\cite{park:2002SCI}, most publications \cite{park:2002SCI,
cho:2002PRB, tsui:2003PRL, d'orazio:2004JMMM} have concentrated on
reporting high $T_\textrm C$ ranging from 116 K \cite{park:2002SCI}
to 285 K \cite{cho:2002PRB} and on interpreting the observed
ferromagnetism in terms of DMS theories \cite{pinto:2005PRB}. It is
only recently that several of the former GeMn reports have been
questioned by structural proofs \cite{kang:2005PRL} and hints
\cite{li:2005PRB} for the formation of intermetallic ferromagnetic
compounds through phase separation in single crystals and
low-temperature molecular beam epitaxy (MBE) fabricated films, respectively. Up to now
only Li \textit {et al.} \cite{li:2005PRB} present - indirect -
indications for the realisation of precipitate free GeMn.
Considering the current discussion on the magnetic properties of
GeMn, a study of the crystal structure, exploring the degree of Mn
dispersion that can be reached in Ge, would obviously be beneficial
for the field.

In this letter we present the first study relating structural
parameters of precipitate free \GEMN\ films to magnetisation data,
providing new insights into the interpretation of the magnetic
properties of GeMn. Although the incorporation of Mn does not induce
explicit phase separation, nanometer sized areas with increased Mn
content compared to the surrounding matrix are detected in
transmission electron microscopy (TEM) analysis. The films show no
overall spontaneous magnetisation down to $2\GRADKM$. The comparison
of the TEM analysis and magnetisation data indicates the formation
of magnetic supermoments due to an inhomogeneous dispersion of Mn in the Ge host matrix. Supermoments appear below a
characteristic ordering temperature, which depends on the volume and
the Mn content of one particular area.

The samples were produced in low-temperature MBE on intrinsic
Ge(001) substrates. After a thermal desorption of the natural oxide,
a flat Ge buffer layer was deposited at a substrate temperature
\TSS\ of $280\GRADCM$. A careful variation of the substrate
temperature during the subsequent co-deposition of nominally 95\% Ge
and 5\% Mn has shown that a film which is free of intermetallic
precipitates was only reproducibly obtained with \TSS\ as low as
$60\GRADCM$ and with a Ge deposition rate of $0.08\ANGS$. The details of
this study will be published elsewhere \cite{ahlers2006}. The
typical film thickness is \NM{200}. Structural analysis of the films
was performed through x-ray diffraction (XRD) and extensive TEM
studies on a JEOL 3010 microscope operating at 300 kV. Special care
was taken during sample preparation to avoid unintentional
post-growth annealing. Chemical analysis was obtained by electron
energy loss spectroscopy (EELS). A commercial superconducting
quantum interference device (SQUID) was used for magnetisation
measurements.

\begin{figure}[tbp]
\includegraphics[width=4.3cm]{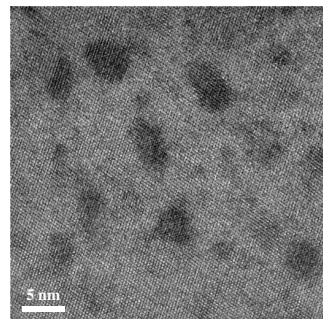}
\caption{\label{fig:HRTEM_TS60} Typical high-resolution TEM
micrograph of \GEMN. Dark contrast reveals cubic clusters which are
coherently bound to the surrounding.}
\end{figure}

No intermetallic precipitates could be detected in XRD for samples with \TS{60}.
Conventional TEM observations on cross-sectional samples (not shown) display perfect epitaxy of the \GEMN\ film on the Ge buffer
layer. The cubic crystalline structure of the buffer layer is
conserved. No dislocation was observed in several TEM micrographs
covering few microns.
\begin{figure}[tbp]
\includegraphics[width=4.3cm]{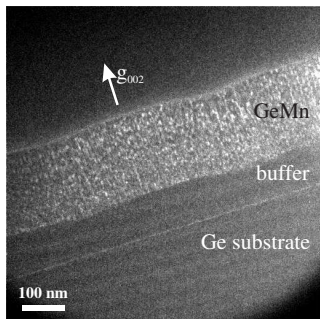}
\caption{\label{fig:TEM_TS60_DF} Dark field TEM overview of a \GEMN\
film fabricated in MBE at $T_\textrm S = 60\GRADCM$. Contrast is
obtained by the selection of the chemically sensitive (002) Ge
reflex. Bright areas correspond to an enhanced Mn content.}
\end{figure}
Furthermore the absence of phase separation
indicates a film free of intermetallic precipitates. Nevertheless, by
selecting the chemically sensitive (002) Ge reflex in dark field
microscopy mode, as depicted in Fig. \ref{fig:TEM_TS60_DF}, a dense
distribution of almost round shaped, nanometer sized areas appeared
in the whole \GEMN\ film. These bright spots represent areas with
increased Mn concentration on substitutional sites of the lattice
compared to the darker matrix, since the (002) reflection is
forbidden in the diamond cubic structure of Ge. The areas will be
denoted as clusters in the following. Their typical diameter is
\NM{4}. EELS measurements with \NM{100} probe size confirm the
stoichiometry of the \GEMN\ alloy for the film. A comparative EELS
analysis of the clusters and the matrix is hampered by their very
small size and dense distribution. An upper limit of 15\% for the
average Mn content on substitutional sites in the clusters is
estimated through the ratio of dark and bright areas supposing that
all the Mn atoms are incorporated into the clusters. Different
levels of brightness of the clusters in the dark field analysis
indicate a slight variation in the Mn content from cluster to cluster. The observations in dark
field TEM are confirmed by high-resolution (HR) TEM micrographs. A typical example is shown in Fig.
\ref{fig:HRTEM_TS60}. The lattice image reveals areas with slightly
darker contrast but still reflecting the same lattice symmetry, that
is, these areas are coherently bound to the surrounding. They are
identified to the clusters shown in Fig. \ref{fig:TEM_TS60_DF} as
they have the same shape and dimensions. Thus the GeMn layer, while
being free of intermetallic precipitates, is composed of a dense
assembly of well defined coherent nanometer sized areas showing
either high or low Mn content.

\begin{figure}[bth]
\includegraphics[width=5cm]{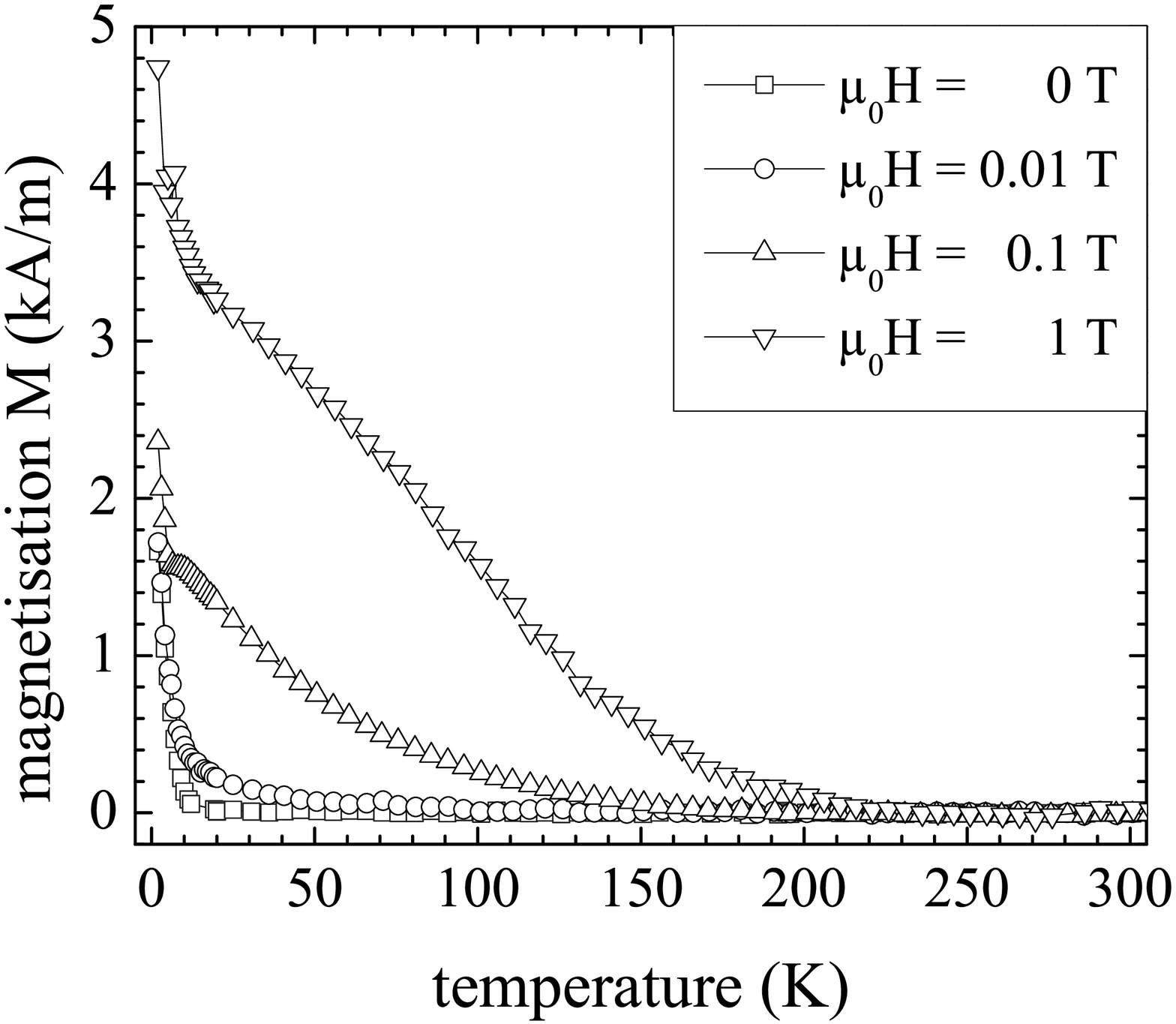}
\caption{\label{fig:SQUID_MT_TS60} Temperature dependence of the
MFC magnetisation of \GEMN~for external fields between $0 \msz
\textrm{T}$ and $1 \msz \textrm{T}$.}
\end{figure}

Temperature dependent magnetisation measurements for different
external magnetic fields are shown in Fig. \ref{fig:SQUID_MT_TS60}.
For all curves the sample was cooled down in the maximum available
field of \MF{7} (maximum field cooled, MFC) and the data recorded
during warm-up in the measurement field. If no external field was
applied during the measurement, the magnetisation decreased from a
finite value at $2\GRADKM$ to zero at about $18\GRADKM$. No
magnetisation at all was detected above this temperature. Nonzero
external magnetic fields in contrast induced non-vanishing
magnetisation values from approximately $18\GRADKM$ up to
approximately $200\GRADKM$ while conserving the steep decrease
between $2\GRADKM$ and $18\GRADKM$. The magnetic properties of the
GeMn layer thus seem to be characterised by two distinct domains
separated at a critical temperature of about $18\GRADKM$. Above this
temperature an overall spontaneous magnetisation of the film can be
excluded.

\begin{figure}[tbp]
\includegraphics[width=8.6cm]{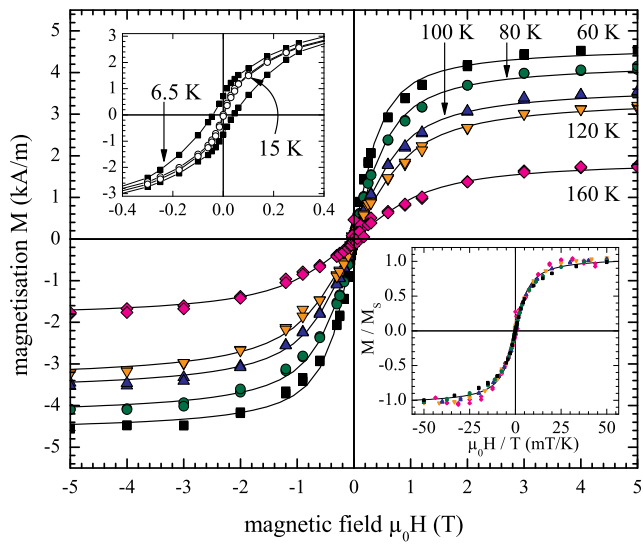}
\caption{\label{fig:SQUID_MH_all_T} (color online) Magnetisation loops taken between $60\GRADKM$ and $160\GRADKM$. Solid lines represent fits
with the Langevin function with $\mu = 435\msz \mu_\textrm B$. (upper inset) Magnetisation loops at $6.5\GRADKM$
and $15\GRADKM$. (lower inset) $M / M_\textrm S$ versus $H / T$. The
symbols represent the same temperature series as in the main figure. The solid line is a Langevin fit with $\mu = 435\msz \mu_\textrm B$.}
\end{figure}

Field-dependent magnetisation loops recorded below and above
$18\GRADKM$ are shown in Fig. \ref{fig:SQUID_MH_all_T}. Again the
sample was cooled down to the measurement temperature in MFC
conditions for all the loops. A hysteresis is observed for
$6.5\GRADKM$, as displayed in the upper inset, with a remanent
magnetisation of $0.73 \msz \textrm{kA/m}$ and a coercive field of
$\mu_0 H_\textrm{c}=41 \msz \textrm{mT}$. The hysteresis gradually
vanishes with increasing measurement temperature towards
approximately $15\GRADKM$. The loops for higher temperatures are
then reminiscent of an atomic paramagnet. But neither does the
high-field saturation of the loops vanish towards higher
temperatures as expected for atomic paramagnets, nor can the loops
be approximated by Brillouin curves using reasonable $g$ and $J$
parameters. They can in turn be described in the infinite limit of
the Brillouin function, that is by a Langevin behaviour, as
indicated by the solid lines in Fig. \ref{fig:SQUID_MH_all_T}, with
\begin{equation} \label{eq:langevin} M(y) = M_\textrm{S}\left(\coth{y} -
\frac{1}{y}\right) \textrm{ , }y = \frac{\mu \mu_0 H}{k_\textrm{B} T}
\end{equation}
where $M$ represents the magnetisation, $M_\textrm{S}$ the
saturation magnetisation of the cluster ensemble and $\mu$ the average magnetic moment per cluster. For each measurement temperature the best Langevin fit is obtained for $\mu = 435\msz \mu_\textrm B$. Moreover, when the same data sets for $T = 60 - 160 \GRADKM$ are plotted as $M$ over $M_\textrm S$ versus external field $H$ over temperature $T$, the data points superimpose as shown in the lower inset of Fig.~\ref{fig:SQUID_MH_all_T}. The Langevin fit of this plot again gives $\mu = 435\msz \mu_\textrm B$. This behaviour is characteristic for superparamagnetism \cite{jacobs1963}. 

Typical clusters have a diameter of
$3-4\msz\textrm{nm}$ and $10-15\msz\%$ of Mn atoms incorporated on
substitutional sites, as identified in TEM. Assuming a spherical shape and an average
moment per Mn atom of $1 - 3\msz \mu_\textrm B$
\cite{li:2005PRB,park:2002SCI,stroppa:2003PRB}, we evaluate a
magnetic moment of about $100 - 600\msz \mu_\textrm B$ per cluster.
The order of magnitude of this estimation is in good agreement with
the average magnetic moment of $\mu=435\msz \mu_\textrm B$ extracted
from the Langevin fits in Fig.~\ref{fig:SQUID_MH_all_T}, while it is
much smaller than the typical moment obtained for intermetallic
precipitates for $T_\textrm S > 60\GRADKM$ \cite{ahlers2006}. The
comparison of the structural and magnetic properties above
$20\GRADKM$ thus indicates that the magnetic moments responsible for
the superparamagnetism are located in the clusters shown in Fig.
\ref{fig:TEM_TS60_DF}. An individual supermoment is then created when a
single cluster turns ferromagnetic below a certain Curie temperature
\TCC. 

The high field saturation magnetisation $M_S$ of the magnetisation loops in Fig. \ref{fig:SQUID_MH_all_T} decreases when the measurement temperature is increased. Since at the same time the average magnetic moments extracted from the Langevin fits stay at a constant level for different measurement temperatures, less clusters seem to contribute to $M_S$. This is equivalent to an increasing amount of clusters that cross their individual \TCC\ and therefore lose their ferromagnetic order. Thus, magnetometry as well as the different Mn contents \cite{dietl:2000S,li:2005PRB} observed in TEM indicate the presence of a distribution of \TCC\ in the cluster ensemble.

As shown in the lower inset of Fig. \ref{fig:SQUID_MH_all_T}, the
system is superparamagnetic up to at least $160\GRADKM$, which
represents a lower limit for the maximum value of \TCC. An upper
limit for this maximum cannot be extracted from this data, mainly
due to a low signal level above $160\GRADKM$. The presence of a nonzero
magnetisation in \MF{1} at $200\GRADKM$ further suggests that the
maximum achievable \TCC\ is larger than $160\GRADKM$. In preliminary
magnetic characterisation the curve form of post-growth annealed
films as well as of ones with Mn content up to 8\% do not differ
from Fig. \ref{fig:SQUID_MT_TS60} and Fig. \ref{fig:SQUID_MH_all_T}.
The obvious difference is a shift of the onset of magnetisation in
nonzero field to values larger than 200 K. This is consistent with
an expected shift of the distribution of supermoments and \TCC\ to
higher values when increasing the Mn concentration.

\begin{figure}[tbp]
\includegraphics[width=8.6cm]{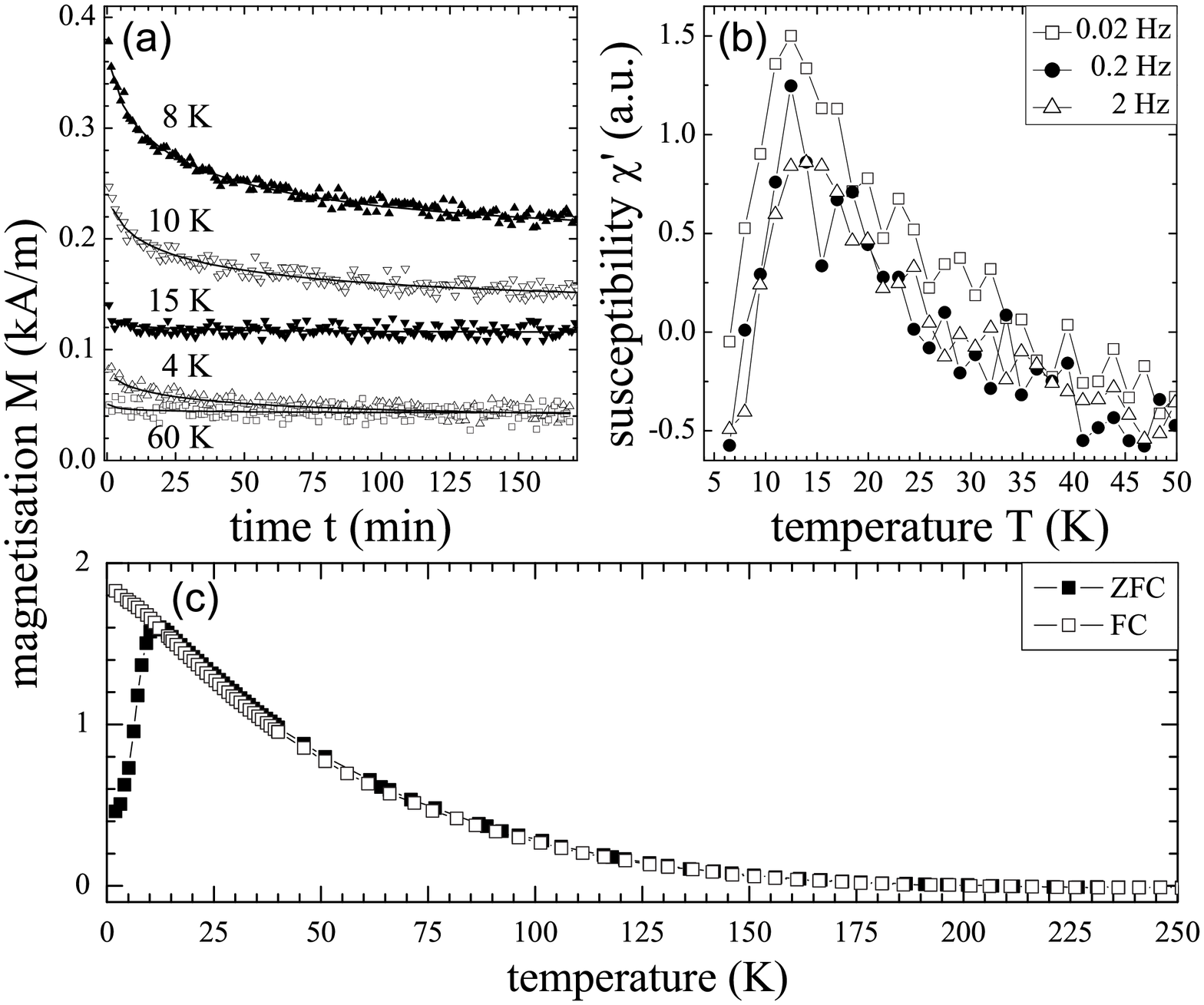}
\caption{\label{fig:SQUID_aging_AC_TS60} (a) Relaxation of the
magnetisation of \GEMN\ after switching off the external field at
different temperatures. Solid lines represent fits to a stretched
exponential decay. (b) Temperature dependent real part of AC
susceptibility measured at different frequencies. (c) FC / ZFC temperature dependent magnetisation. \MF{0.1}.}
\end{figure}

Fig. \ref{fig:SQUID_MT_TS60} and Fig. \ref{fig:SQUID_MH_all_T} at
first glance seem to indicate the presence of a spontaneous
magnetisation below a critical temperature of approximately
$18\GRADKM$. However, the magnetic properties below $18 \GRADKM$
depend on the sample history. This can be observed in relaxation
measurements for zero field cooled (ZFC) samples in Fig.
\ref{fig:SQUID_aging_AC_TS60}(a), where magnetisation is measured
versus time after switch-off of an externally applied field. There
is an obvious relaxation of the magnetisation for temperatures below
$15 \GRADKM$. Above $15\GRADKM$, relaxation is strongly suppressed.
The solid lines in Fig. \ref{fig:SQUID_aging_AC_TS60}(a) represent
fits to a stretched exponential decay $\sim
\exp\left[-(t/\tau)^{1-n}\right]$ with $n$ values between $0.5$ and
$0.6$. In Fig. \ref{fig:SQUID_aging_AC_TS60}(b), a peak in the real
part of the low field AC susceptibility measured at different
measurement frequencies is observed at temperatures around $14\GRADKM$. This susceptibility peak changes slightly
both in height and position depending on the measurement frequency.
The relative shift of the peak position per frequency decade is
$0.03$. We interpret the measurements in terms of a metastable
state. This metastable state can be induced by a multi-valley
structure in the free energy phase space of the system at low
temperatures, resulting in magnetisation relaxation phenomena that
become visible in ageing effects and in a peak shift of the AC
susceptibility \cite{mydosh1993}. Further indications supporting
this interpretation are the observation of a nonreversibility of
magnetisation at $18\GRADKM$ in Fig.~\ref{fig:SQUID_aging_AC_TS60}(c), that is a bifurcation of field cooled (FC) and ZFC
temperature dependent magnetisation measurements, and a peak  at $18\GRADKM$ in the ZFC curves. Furthermore, measurements taken during zero field cooldown lead to a zero magnetic signal (not shown). A multi-valley structure of the free energy
landscape below $15\GRADKM$ can be due to the existence of a
glass-like state or interacting blocked particles
\cite{dormann:1999JMMM}. Neither of the two possibilities can be
excluded from our measurements. Both of them would induce the steep
decrease in magnetisation between $2 \GRADKM$ and $18 \GRADKM$ in
Fig. \ref{fig:SQUID_MT_TS60} as well as the gradually vanishing
hysteresis in Fig. \ref{fig:SQUID_MH_all_T} in the same temperature
range. Nevertheless, the clear indications for metastability make a
transition to a ferromagnetic state highly unlikely. Furthermore,
temperature dependent resistivity measurements of our samples reveal
typical insulating behaviour with a diverging resistivity towards
zero temperature. Irregularities in the temperature dependent
resistivity curve are expected in case of magnetic ordering
\cite{li:2005PRB,kaminski:2003PRB,nazmul:2003PRB}. No irregularity and therefore no hint for a magnetic
percolation is observed in the temperature range where the system
undergoes a transition between the superparamagnetic to a metastable
regime in our samples. Thus from our experiments, suggesting well
defined and stable clusters with enhanced Mn concentration on
substitutional Ge lattice sites, we cannot confirm a ferromagnetic
percolation \cite{li:2005PRB} conjectured to result from magnetic
"spin clusters" which grow in size with decreasing temperature. We
suggest that the ferromagnetism of a cluster is mediated by holes
which are spatially localised inside this cluster
\cite{kaminski:2003PRB, jaroszynski2005}.

To conclude, we have shown that as-grown \GEMN\ layers free of
intermetallic precipitates, produced by low temperature MBE at
\TS{60} are very close to the Mn dispersion level which can be
realistically expected for this kind of extreme epitaxy conditions.
We report the first observation of Mn enriched clusters with an
upper limit of 15\% Mn per cluster. The
magnetic properties of GeMn films are clearly a consequence of the
dense ensemble of nanometer sized clusters. Each of these clusters
turns ferromagnetic below its own characteristic transition
temperature \TCC\ which depends on its Mn content and its volume.
The ferromagnetism of a cluster is mediated through holes which can
freely move only inside this cluster. The surrounding matrix in
contrast seems to lack freely moving holes. The clusters then carry
individual magnetic supermoments which lead to the observation of
superparamagnetism above approximately $18\GRADKM$ and a metastable
state at lower temperatures. The physical origins of the metastable state have to be further investigated. The lower limit of the maximal
$T_\textrm C ^\textrm{Cluster} = 160\GRADKM$ confirms the potential
of a Ge based DMS for spintronic applications.
\begin{acknowledgments}
This work was supported by Deutsche Forschungsgemeinschaft via SFB
631. The authors gratefully acknowledge access to the SQUID magnetometer at the
Walther-Meissner-Institut, Garching, as well as discussions with R. Gross, M. Opel,
C. Jäger, C. Bihler and M. S. Brandt.
\end{acknowledgments}
\paragraph{Note added. \textemdash } After this manuscript was sumbmitted, we learned of a related work \cite{jamet:2006NM} reporting the observation of GeMn nanocolumns.


\end{document}